\documentclass[aps, nobibnotes ,secnumarabic ,showpacs ,amssymb, amsmath, prb
]{revtex4}

\usepackage{epsfig}
\usepackage{endnotes}
\let\footnote=\endnote

\begin{document}

\def\p{\phi}\def\P{\Phi}\def\a{\alpha}\def\e{\epsilon}
\def\be{\begin{equation}}\def\ee{\end{equation}}\def\l{\label}
\def\0{\setcounter{equation}{0}}\def\b{\beta}\def\S{\Sigma}\def\C{\cite}
\def\r{\ref}\def\ba{\begin{eqnarray}}\def\ea{\end{eqnarray}}
\def\n{\nonumber}\def\R{\rho}\def\X{\Xi}\def\x{\xi}\def\la{\lambda}
\def\d{\delta}\def\s{\sigma}\def\f{\frac}\def\D{\Delta}\def\pa{\partial}
\def\Th{\Theta}\def\o{\omega}\def\O{\Omega}\def\th{\theta}\def\ga{\gamma}
\def\Ga{\Gamma}\def\t{\times}\def\h{\hat}\def\rar{\rightarrow}
\def\vp{\varphi}\def\inf{\infty}\def\le{\left}\def\ri{\right}
\def\foot{\footnote}\def\vep{\varepsilon}\def\N{\bar{n}(s)}
\def\k{\kappa}\def\sq{\sqrt{s}}\def\bx{{\mathbf x}}\def\La{\Lambda}
\def\bb{{\bf b}}\def\bq{{\bf q}}\def\cp{{\cal P}}\def\tg{\tilde{g}}
\def\cf{{\cal F}}\def\bN{{\bf N}}\def\Re{{\rm Re}}\def\Im{{\rm Im}}
\def\bk{\hat{\mathbf{k}}}\def\cl{{\cal L}}\def\cs{{\cal S}}\def\cn{{\cal N}}
\def\cg{{\cal G}}\def\q{\eta}\def\ct{{\cal T}}\def\bbs{\mathbb{S}}
\def\bU{{\mathbf U}}\def\bE{\hat{\mathbf e}}\def\bc{{\mathbf C}}
\def\vs{\varsigma}\def\cg{{\cal G}}\def\ch{{\cal H}}\def\df{\d/\d }
\def\mz{\mathbb{Z}}\def\ms{\mathbb{S}}\def\kb{\hat{\mathbb
K}}\def\cd{\mathcal D}\def\mj{\mathbf{J}}\def\Tr{{\rm Tr}}
\def\bu{{\mathbf u}}\def\by{{\mathrm y}}\def\bp{{\mathbf p}}
\def\k{\kappa} \def\cz{{\mathcal Z}}\def\ma{\mathbf{A}}
\def\me{\mathbf{E}}\def\ra{\mathrm{A}}
\def\ru{\mathrm{u}}\def\rP{\mathrm{P}}\def\rp{\mathrm{p}}\def\z{\zeta}
\def\my{\mathbf Y}\def\ve{\varepsilon}\def\bw{\mathbf
W}\def\hp{\hat{\p}}\def\hh{\hat{h}}\def\hx{\hat{\x}}\def\hk
{\hat{\kappa}}\def\hj{\hat{j}}\def\eb{\mathbf{e}}\def\bj{\mathbf{j}}
\def\tO{\tilde\o}\def\cS{{\cal
S}}\def\os{\overline{\s}}\def\K{\hat{K}}\def\mH{\mathcal{H}}
\def\to{\rightarrow}\def\t{\times}\def\ola{\overline{\la}}
\def\xb{\bar{\xi}}\def\qb{\bar{\eta}}\def\by{\bold{y}}\def
\bq{\mathbf q}

\title{ Particle Production in the Field Theories with Symmetry}
\author{J.~Manjavidze}\email{joseph@jinr.ru}
\affiliation{ Andronikashvili Institute of Physics, Tbilisi State
University, Georgia,} \affiliation{Joint Institute for Nuclear
Research, Russia}

\begin{abstract}

The field theory with high space-time symmetry is considered
with the aim to examine the mass-shell particles production
processes. The general conclusion is following: no real
particle production exists if the space-time symmetry
constraints are taken into account. This result does not depend
on the concrete structure of Lagrangian.

\end{abstract}


\received{}

\maketitle

\section{Introduction}\0

The purpose of present article is to calculate the cross section of
inelastic processes in theories with high space-time symmetry. We
will consider a case when the action, $S$, have the nontrivial
extremum at $u(x)$, \be u(x): \f{d S(\vp)}{\d\vp(x)}=0,\l{equ}\ee
where $\vp$ is the boson field. The quantitative consideration of
this question seems important since although there exists a number of
approaches to the canonical quantum field theory formalism in the
vicinity of extended field $u(x)\neq0$, see e.g. \C{goldst, jakiw},
the observables practically were not considered because of the
complicated problem with symmetry constraints \footnote{The important
question of symmetry constraints was considered in \C{diracc}.}. A
theory in which the consequences of broken by $u(x)$ symmetry is
taken into account explicitly will be called as "the field theory
with symmetry" understanding that $u(x)$ is the result, at least, of
high space-time symmetry of action $S(\vp)$.

The main physical result of present paper looks as follows: the
transition of interacting field into the mass-shell particles state,
and vice versa, is impossible in the field theories with symmetry. We
will consider the general case, $u(x)$ is not necessarily the soliton
field which is absolutely stable against decay on the particles, see
e.g. \C{jakiw}. The introduction into the necessary formalism and
quantitative prove of $2d$ solitons stability against particles decay
was described in the review paper \C{prepr}. The main formal result
of this work is the further development of formalism \C{prepr, tmf}
which is able to solve particle production problem in the $4d$ field
theories with symmetry.

It will be shown explicitly at the very end that the $m-$ into
$n-$particles transition cross section times a flux factor,
$\R_{mn}$, is trivial: \be \R_{mn}=0,~~ \forall (m,n)>0, \l{o2}\ee if
the field theory with symmetry is considered.

In Sec.2 the cross section $\R_{mn}$ will be introduced and in Sec.3
the method of calculation of $\R_{mn}$ will be described. The prove
of Eq. (\r{o2}) will be given in Sec.4. A short list of unsolved
problem will be given in the last Sec.5.

The conclusion (\r{o2}) can be extended directly on the gluon
production case in non-Abelian gauge theory without matter
(quark) fields, see also Sec.5.

\section{Integral representation for generating functional of
$\R_{mn}$}

It will be seen that the used formalism allows to act {\it ex
adverso}. So, we will introduce $S$-matrix using ordinary LSZ
reduction formalism. The conclusion (\r{o2}) is general, it does not
depend on the concrete form of theory Lagrangian, $L$. For this
reason one can have in mind the simplest $4d$ conformal scalar field
theory: \be L=\f{1}{2}(\pa\vp)^2- \f{g}{4}\vp^4,~g>0, \l{lagr}\ee as
an example to consider massless scalar particles production, see
Appendix A where particles production in the theory with Lagrangian
(\r{lagr}) is described.

\subsection{\it Generating functional}

We expect that the interaction of internal states with external
particles is switched on adiabatically, i.e. that the external
fields can not have an influence to the spectrum of interacting
field perturbations in the case of field theory with symmetry.
The builded formalism will correspond to this basic condition.

So, the $(m+ n)$-point Green function $G_{mn}$ is defined by a
formulae: $$ G_{mn}(y_1,y_2,...,y_m; x_1,x_2,...,x_n)=\int
D\vp\prod_{k=1}^m \vp(y_k)\prod_{k=1}^n \vp(x_k) e^{iS(\vp)}.$$ The
LSZ reduction formula means that the external legs (massless
particles in the considered case) must be amputated, i.e. the
amplitude is defined by the expression \C{caruzerss}, see also
\C{physrep}:
$$A_{mn}(y_1,...,y_m;x_1,...,x_n)=\int
D\vp\prod_{k=1}^m\pa^2_{y_k}\vp(y_k)\prod_{k=1}^n
\pa^2_{x_k}\vp(x_k) e^{iS(\vp)},$$ and the amplitude in the
energy-momentum representation looks as follows, see Fig.1: \be
a_{mn}(q_1,....q_m ;p_1,p_2,...,p_n)=\int D\vp\prod_{k=1}^m
\Ga(q_k;\vp)\prod_{k=1}^n \Ga^*(p_k;\vp) e^{iS(\vp)},
\l{2.2a1}\ee where \be \Ga(q;\vp)=\int dx e^{-ixq}
\pa^2\vp(x),~q^2=0, \l{2.2a2}\ee is the external particles
annihilation vertex. It must be noted absence of the
energy-momentum conservation $\d$-functions in the definition
of the amplitude (\r{2.2a1}). Considering the extended field
configurations, $u(x)$, the conservation of the external
particles energy and momentum is the isolated problem, see
\C{tmf}. In considered case this question is not important.

\begin{figure}[ht]
\begin{center}
\begin{tabular}{c}
\mbox{\epsfig{file=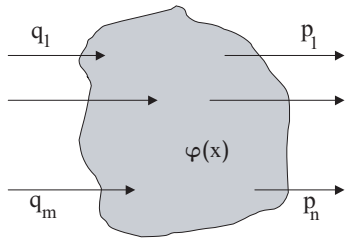, width=0.25\textwidth}}
\end{tabular}
\end{center}
\caption{\it\footnotesize The amplitude $a_{mn}$. The plane wave
$e^{-iq_kx_k}$, $k=1,2,..., m$, is associated to each in-coming
particle and $e^{ip_ky_k}$, $k=1,2,...,n$, to each out-going one. It
is supposed that $q_k^2=p_k^2=0$. The integration over $\vp(x)$ must
be performed.}
\end{figure}

The common point of view on the multiple production gives the
method of generating functionals, $R(z)$, through the
expression:
$$R(z)=\sum_{m,n}\f{1}{m!n!}\int
d\o_m(z,q)d\o_n^*(z,p)|a_{mn}(q_1,...,q_m;p_1,..., p_m)|^2,$$ see
Fig.2, where the usual probe function $z(q)$ was introduced:
$$d\o_m(z,q)=\prod_{k=1}^m\f{d^3q_k z(q_k)} {(2\pi)^3
2\ve(q_k)},~~\ve(q)=\sqrt{q^2}$$ and $dz/dz^*\equiv0$. As a result,
\be R(z)=\sum_{m,n=0}^{\infty}\f{1}{m!n!}\int
D\vp^+D\vp^-\le\{e^{iS(\vp^+)}
{N(z,\vp^\pm)}^m\ri\}\le\{e^{-iS^*(\vp^-)}{N^*(z,\vp^\pm)}^n\ri\},\l{2.4b}\ee
where \be N(z,\vp^\pm)=\int d\o_1(z,q)
\Ga(q;\vp^+)\Ga^*(q;\vp^-)\l{N}. \ee

\begin{figure}[!ht]
\begin{center}
\begin{tabular}{c}
\mbox{\epsfig{file=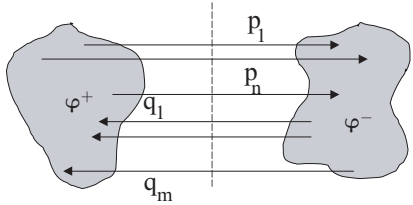, width=0.3\textwidth}}
\end{tabular}
\end{center}
\caption{\it\footnotesize The diagram for $(m+n)$-particle
absorption part of
vacuum-into-vacuum amplitude. Each in-coming
line carry the factor $z(q_k)$, $k=1,2,..., m$, and
$z^*(p_k)$, $k=1,2,..., n$ is associated with
out-going one. The field
$\vp_+$ is defined on the Mills complex time contour $C^+$
and $\vp_-$ is defined on $C^-=C^{+*}$ \C{prepr}. The
summation over $m$ and $n$ and the independent integration
over $\vp_+$ and $\vp_-$ must be performed. The vertical dotted
line cross mass-shell particle lines}
\end{figure}

The quantity $R(z)|_{z=1}$ coincides with the imaginary part of the
vacuum-into-vacuum transition amplitude, see Fig.2. In turn,
$R(z)|_{z=0}=|a_{00}|^2$ is the modulo squire of vacuum-into-vacuum
transition amplitude. Correspondingly the unnormalized cross section
of $(2\to n)$ particle transition is equal to
$$\R_{2n}=\prod^2_{i=1}(2\pi)^32\ve(q_i) \f{\d}{\d z(q_i)}
\prod^n_{i=1}(2\pi)^32\ve(p_i) \f{\d}{\d z^*(p_i)}R(z)|_{z=z^*=0} .$$
The correlation functions are defined through variation of $\ln R$
over $z$. The inclusive cross sections are defined by variation of
$R(z)$ over $z^*$ at $z^*=1$.

\subsection{\it Dirac measure}

We will use following following representation for $R(z)$
\C{prepr, physrep}: \be
R(z)=\lim_{j=e=0}e^{-i\kb(je)}\sum_{m,n=0}^{\infty}\f{1}{m!n!}\int
DM(\vp)e^{iU(\vp,e)/\hbar} {N(z;\vp)}^m{N^*(z;\vp)}^n.
\l{2.13ax}\ee It must be underlined that the representation
(\r{2.13ax}) means calculation of the r.h. part of depicted on
Fig.3 diagram.

The operator \be 2\kb(je)=\Re\int_{C^+} dx\f{\d}{\d
j(x)}\f{\d}{\d e(x)}\l{2.6b}\ee generates quantum excitations
of the field $\vp(x)$, where $C^+$ is the Mills time contour
\C{prepr}: \be C^+: t\to t+i\ve, ~\ve\to+0.\l{mill}\ee The
auxiliary variables $e(x)$ and $j(x)$ must be taken equal to
zero at the very end of calculations. We will assume that, for
example, \be \f{\d e(\bx,t)}{\d
e(\bx',t')}=\d(\bx-\bx')\d(t-t')\l{presc}\ee iff $(t,t')\in
C^+$. Otherwise this derivative is equal to zero identically.
The functional: \be U(\vp,e)=S(\vp+e)-S(\vp-e)-2\Re \int_{C_+}
dx e(x)\f{\d S(\vp)}{\d\vp(x)} \l{U}\ee describes the
interactions in a given field theory. It is not hard to see
that for example \be U(\vp,e)= g\Re\int_{C_+} dx
e^3(x)\vp(x)\l{Uex}\ee for $g\vp^4$ theory. At last $DM$ is the
(Dirac or $\d$-like) differential measure: \be DM(\vp)=\prod_x
d\vp(x)\d\le(\f{\d S(\vp)}{\d\vp(x)}+\hbar j(x)\ri)
.\l{2.6aa}\ee Performing calculations one must take into
account the prescription (\r{presc}). Actually the arguments of
$DM(\vp)$ and $U(\vp,e)$ are defined on the whole contour
$C=C^++C^-=C^+-C^{+*}$ \C{prepr}.

A few words in connection with qualitative meaning of representation
(\r{2.13ax}). The representation (\r{2.13ax}) can be derived from
(\r{2.4b}) extracting from the fields $\vp^\pm$ the "mean" field
$\vp(x)$ and $e(x)$ is the deviation from it, $\vp^\pm(x)=\vp(x)\pm
e(x)$, with boundary condition: \be e(x\in \s_{\infty})=0,
\l{boun}\ee see Fig.4. The integration over $e(x)$ gives functional
$\d$-function of Eq. (\r{2.6aa}) \C{prepr}. The source $j(x)$ was
introduced to take into account the non-linear terms over $e(x)$,
i.e. the variation over $j(x)$ generates the quantum corrections.

Notice absence of $e(x)$ in the argument of $N(z; \vp)$ because
of the prescription (\r{boun}) and since $\Ga(q;\vp)$ is
accumulated at $\s_\infty$ if $q^2=0$ in the theories with
symmetry. Correspondingly there is not an influence of external
state, which is labeled by $z(q)$, on the argument of
$\d$-function in (\r{2.6aa}). Thus produced particles state
does not have an influence on the internal fields spectrum. We
will return to this question later.

\begin{figure}[!ht]
\begin{center}
\begin{tabular}{c}
\mbox{\epsfig{file=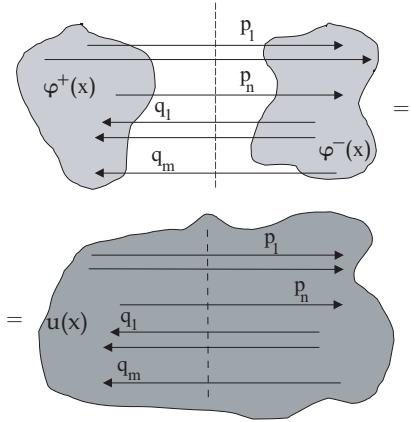, width=0.3\textwidth}}
\end{tabular}
\end{center}\vskip -0.5cm
\caption{\it\footnotesize Optical theorem. Summation over $m$
and $n$ is assumed. The contributions of r.h.s. diagram are counted by the
coordinates $(\ga)$ of factor space $W$.}
\end{figure}

So, we restrict ourself by the direct calculation of the observable
quantity, $\R_{mn}$. This is crucial since allows to take into
account the consequences of non-measurability of the quantum phase of
amplitude $a_{mn}$ which is canceled in $\R_{mn}$. Practically it is
the additional for quantum systems dynamical principle of time
reversibility, see the comment to Fig.4 and \C{GCPr}. It means that
all acting in the system forces must compensate each other {\it
strictly} in the frame of condition (\r{boun}) \footnote{ This
reminds the principle of d'Alembert.}, i.e. in the quantum case we
have new equation of motion instead of (\r{equ}), see (\r{2.6aa}):
\be \f{\d S(\vp)}{\d \vp(\bx,t)}=-\hbar j(\bx,t)\l{ab}\ee if
interaction with external field is switched on adiabatically. In
opposite case the $z$-dependent term appears in the r.h.s. of Eq.
(\r{ab}). The source (force) $j(\bx,t)$ in Eq. (\r{ab}) generates
quantum excitations. We will search the solutions of Eq. (\r{ab})
expanding them in vicinity $j=0$.

A short qualitative description of corresponding to (\r{ab}) {\it
generalized corresponding principle} (GCP) one can find in \C{GCPr},
the detailed derivation of Eq. (\r{ab}) is given in the review paper
\C{prepr}. Notice also that (\r{ab}) is reduced to the correspondence
principle of Bohr in the limit $\hbar=0$. GCP means that the
contributions into functional integral for $\R_{mn}$ are defined by
the complete set of solutions of {\it strict} equation (\r{ab})
\C{prepr}. This is why we can act {\it ex adverso}: Eq. (\r{ab})
defines {\it all} necessary and sufficient real time contributions
into $\R_{mn}$.

\begin{figure}[!ht]
\begin{center}
\begin{tabular}{c}
\mbox{\epsfig{file=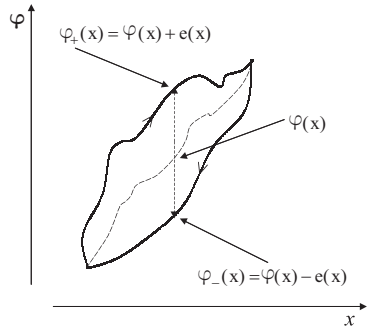, width=0.3\textwidth}}
\end{tabular}
\end{center}\vskip -0.5cm
\caption{\it\footnotesize The dynamics along the mean trajectory $\vp(x)$ is time
reversible since the total action $S(\vp_+)-S^*(\vp_-)=
S(\vp_+)+S(\vp_-)$ describes closed path motion, $<in|out><in|out>^*=
<in|out><out|in>$,
in the frame of boundary condition (\r{boun}).}
\end{figure}

As the result, to find $R(z)$ in the frame of ordinary
canonical scheme, see e.g. \C{goldst, jakiw, fadde}

(I) one must start from the equation: \be \f{\d
S(\vp)}{\d\vp(x)}=0,\l{2.17}\ee see the GCP representation
(\r{2.13ax}). Having the solution $u(x)$ of this equation\\
(II) one can find $u_j(x;u)$ from the complete equation: \be
u_j:~\f{\d S(\vp)}{\d\vp(x)}= j(x),~u_j(x)|_{j=0}=u(x),\l{2.18}\ee in
the form of the series over $j(x)$. It is evident that in that case
we describe perturbations in the vicinity of $u(x)$. The same problem
is solved in the stationary phase method. This step is reduced
practically to the search of the particles propagator in the
"external field" $u(x)$. Generally this problem is unsolvable since
in this case the 4-momentum is not conserved along particles
trajectory. It must be noted that the expansion of $u_j$ over $j$
leads to the expansion over positive powers of interaction constant,
i.e. presents the "weak-coupling" expansion.

(III) Next, \be \R_{mn}=e^{-i\kb(j,e)} \sum_{\{u_c\}} e^{iU(u_j,e)}
{N(u_j)}^m{N^*(u_j)}^n \det (u_j)^{-1}, \l{2.8}\ee where $\det(u_j)$
is the functional determinant:
$$\det(u_j)^{-1}=\int \prod_x d\vp(x)\d\le(\f{\d^2 S(u_j)}{\d
u_j(x)^2}\vp(x)\ri).$$

(IV) The last step is the calculation of the perturbation
series generated by the operator $\kb$. Partial cancelation of
contributions, which leads to the $\d$-like measure
(\r{2.6aa}), unchange the convergence radii. It can be shown
that the obtained perturbation series has zero convergence
radii \C{prepr}.

It is not hard to see that the naive use of solution of Eq.
(\r{2.18}) leads to $\Ga(q;u_j)\neq0$.

\section{Theories with symmetry}

The crucial point of our analysis is the observation that $\Ga(q;u)$
stocks up on the remote hypersurface $\s_\infty$ and that the
external particle belongs to mass shell, $q^2=0$. Indeed, the vertex
$\Ga$ can be rewritten identically in the form: \be \Ga(q;\vp)=\int
dx \pa_\mu\le((\pa^\mu +2iq^\mu)e^{-iqx}\vp(x)\ri) \l{ga1}\ee if
$q^2=0$ and $\vp$ is the nonsingular function.

Our aim is to investigate $\Ga(q;u)$, where $u(x)$ is the solution of
Eq. (\r{2.17}), in all orders over $\hbar$ in the frame of the
condition that the energy of $u(x)$ is finite: \be H(u)= \int
d\bx\le(\f{1}{2}(\pa_t u)^2+\f{1}{2}(\nabla u)^2+v(u)\ri)<\infty.
\l{acti}\ee  It will be shown that $\Ga(q;u)=0$ since actually $u(x)$
is the nonsingular well localized object even in quantum case, i.e.
(\r{acti}) is rightful in all orders of $\hbar$.

The way of calculation of $\R_{mn}$ shown at the end of
previous Section is quite cumbersome. Moreover, the effect of
"symmetry breaking by $u(\bx,t)$" is hidden in this approach,
there is no obvious way to find the consequences of symmetry
constraints. That is why the another way of computation of
integral (\r{2.13ax}) was chosen.

Having a theory on  $\d$-like measure (\r{2.6aa}) one may adopt such
most powerful method of classical mechanics as the transformation of
variables, see also  \C{prepr}. This is the one of important
consequences of $\d$-likeness of measure (\r{2.6aa}). We will
consider in present paper the transformation, \be u:~
\vp(\bx,t)\to~\ga(t)\in W, \l{1.3'}\ee where the new finite set of
"fields", $\{\ga\}$, are the functions only of the time, $t$, see
also Fig.3. Noting that $\vp(\bx,t)$ is the function of
$continuous~set$ of variable $\bx$ and that the new "fields"
$\ga_i(t)$ is labeled by the $countable~set$ of the indexes,
$i=1,2,...,\nu,$ the mapping (\r{1.3'}) means infinite reduction.
This reduction of degrees of freedom is the main formal problem
considered in details in present Section, see also \C{prepr}.

It must be noted also that the used formalism is Lorenz
non-covariant. For this reason we will distinguish space and
time components, $x=(\bx,t)$. This circumstance is not crucial
since we calculate the cross section, $\R_{mn}$, which always
is defined in the definite Lorenz frame. We will explain in the
Appendix B why the general case $\ga=\ga(\bx,t)$ have not been
realized. So, we will not pay attention during subsequent
calculations to the space components, $\bx$, since they are the
insufficient variables. Actually $u(\bx,t)$ would be the
singular distribution function of time because of the quantum
perturbations \C{tmf} but we will see that the singularities of
$u(\bx,t)$ are integrable and do not gives an influence on the
final result (\r{o2}).

The transformation (\r{1.3'}) is generated by a strict
solution, $u(\bx,t;\ga_0)$, of the Lagrange equation (\r{equ})
where $\{\ga_0\}$ is the set of integration constants. The
complete set of solutions of nonlinear $4d$ equations of motion
like (\r{equ}) is unknown\footnote{The list of known solutions
of Eq. (\r{equ}) is given e.g. in \C{actorr}.} and we are
forced to assume that the classical field, $u(\bx,t;\ga_0)$, of
necessary property (\r{acti}) exists. This is a main lead-in
assumption of the approach, the explicit form of
$u(\bx,t;\ga_0)$ will not be important for us.

The set of new fields, $\{\ga(t)\}$, will be defined by the set
$\{\ga_0\}$, i.e. we will describe quantum dynamics mapping the
problem with symmetry into the {\it factor space} $W$,
$$\{\ga(t)\}_{t=0}=\{\ga_0\}\in W.$$ The approach goes back to the old idea of
statistical systems description in terms of {\it collective
variables} \C{bogol} \footnote{For example, $\ga_0$ may define the
space-time position of $u$-th maximum, its scale, etc. In other words
the set $\{\ga_0\}$ defines the integral, i.e. the "collective", form
of $u(\bx,t)$. Allowing for the symmetry constraints only the
collective variables remain free \C{takhta}.}. The simple explanation
of topology side of the transformation (\r{1.3'}) was described in
the transparent papers \C{smalee}, see also textbook \C{arnoldd} and
\C{marsden}. The paper \C{fadde} is also useful since clarify
Hamilton description to the extended, soliton, field configurations,
farther details one can find in \C{takhta}. One can note also
existence of the suggestion \C{nohl} to leave the frame of canonical
schemes to quantize the extended fields.

Actually our approach to the quantum field theory with symmetry
consists from two parts. First one stands of the mapping into
$W=G/G_w$, where $G_w\in G$ is the symmetry group of $u$
\footnote{This explains why the definition: "field theory with
symmetry" was introduced. For example, $W=O(4,2)/O(4)\t O(2)$ in the
conformal field theory with symmetry if $u$ is the $O(4)\t
O(2)$-invariant solution of Eq. (\r{equ}) \C{dealf}.}. The second one
contains dynamics, see also \C{fadde}. The problem of quantization
comes into existence only in the second part.

One can call following useful geometrical interpretation of the
"collective variables" approach \C{prepr}. The set of
parameters $\{\ga_0\}$ form the factor space $W$ and
$u(\bx,t;\ga_0)$ belongs to it {\it completely}. The mapping of
dynamics into $W$ form in it the finite-dimensional
hypersurface. For example, the hypersurface compactify into the
Arnold's hypertorus if the classical system is completely
integrable, see additional references in \C{arnoldd} and
\C{takhta}. Then half of parameters $\ga$ are the radii of the
hypertorus and other half are the angles.

Description of quantum system in terms of the collective-like
variables $\ga(t)$ means the description of random deformations of
such hypersurface, i.e. of the surface of Arnold's hypertorus in the
case of completely integrable system. That is why our approach
describes just the fluctuations of $u(\bx;\ga(t))$ at the expense of
fluctuations of $\ga(t)$. Therefore, our formalism describes {\it the
fluctuations of $u(x)$}, instead of usually considered canonical
formalism which describes the fluctuations in {\it the vicinity} of
$u(x)$,  Sec.2.2.

Therefore, the main step of our calculations is the reduction of
field-theoretical problem with symmetry to the quantum-mechanical
one, where $(\x(t),\q(t))\in\{\ga(t)\}$ are the generalized
coordinates and momenta of the particle which is moving in $W$. It
should be noted that in the frame of to-day knowledge it is
impossible to present the complete set of first integrals of motion
in involution considering the equations of type (\r{equ}).
Nevertheless we incline to interpret the reduction of degrees of
freedom (\r{1.3'}) as the consequence of symmetry constraints
\footnote{ One can find the example of analogous reduction in the
simplest completely integrable system in \C{prepr}. Our
interpretation of the reduction is rightful in that case, see also
\C{takhta}.}.

It is evident that being the infrared stable the quantum-mechanical
perturbations of $\ga(t)$ unchange the conclusion that $u(x)$ is the
well localized field, i.e. $u(x)|_{x\in\s_\infty}=0,~\forall\hbar$.
That is why we come to (\r{o2}) in all orders over $\hbar$.

\subsection{\it Mapping into $T^*W$}

The method of transformation (\r{1.3'}) looks as follows
\C{prepr}. One can simplify calculations considering the case
of $N(z;\vp)=1$ since interactions with external fields are
switched on adiabatically. Then we have: \be
\R_0=\lim_{j=e=0}e^{-i\kb(j,e)}\int DM e^{iU(\vp,e)},
\l{3.35}\ee where $U(\vp,e)$ is the odd functional over $e$ and
\be 2\kb(j,e)=\Re \int_{C^+} d\bx dt\f{\pa}{\pa
j(\bx,t)}\f{\pa}{\pa e(\bx,t)} ,\l{3.36}\ee \be
DM=\prod_{\bx,t}
d\vp(\bx,t)\d(\pa_\mu^2\vp(\bx,t)+v'(\vp)-j(\bx,t)).
\l{3.37}\ee One may shift Mills time contour $C^+$, see
(\r{2.6b}), on the real time axis since the description of
fluctuations of $u(\bx;\ga)$ in terms of $\ga(t)$ would be free
from light-cone singularities \C{prepr}. This slightly
simplifies calculations but the analytical continuation on the
real time axis should be done carefully if $u(x)$ have
nontrivial topology \C{tmf}, see also \C{smalee}.

The distinction of field theory from quantum mechanics consists in
the presence of the space degrees of freedom. To look into this
problem let us consider the formalism on the "smoothed"
$\d$-function: \be \lim_{\e=0}\d_\e(x)=\d(x).\l{del}\ee It obeys the
property of the ordinary $\d$-function: $$ \int dx
f(x)\d_\e(x-a)=f(a)(1+O(\e)), ~\e\to 0.$$ At the same time $\d_\e(0)$
is finite,
$$\d_\e(0)=O(1/\e).$$

Then, introducing the $auxiliary$ variable $\pi(\bx,t)$: \be 1=\int
\prod_{\bx,t} d\pi(\bx,t) \d(\pi(\bx,t)-\dot\vp(\bx,t)) \l{3.38}\ee
we come to the measure: \be DM=\lim_{\e=0}\prod_{\bx,t}
d\vp(\bx,t)d\pi(\bx,t)\d_\e\le(\dot\vp(\bx,t)-\f{\d H_j(\pi,\vp)}{\d
\pi(\bx,t)}\ri) \d_\e\le(\dot\pi(\bx,t)+\f{\d H_j(\pi,\vp)}{\d
\vp(\bx,t)}\ri). \l{3.39}\ee  The Hamiltonian looks as follows: \be
H_j=\int d\bx\le\{\f{1}{2}\pi^2(\bx,t)+\f{1}{2}(\nabla\vp(\bx,t))^2+
v(\vp)-j(\bx,t)\vp(\bx,t)\ri\}.\l{ham}\ee The independency of $\vp$
and $\pi$ is not important for us. Introduction of the auxiliary
variable $\pi$ is useful since in this case we come to the first
order formalism and this will be important.

Let us introduce the unit: \be 1=\f{1}{\D_\e}\int
\prod_{i=1}^\nu\prod_{t} d\x_i(t)d\q_i(t)\prod_{\bx,t}
\d_\e(\vp(\bx,t)-u(\bx;\x,\q))\d_\e(\pi(\bx,t)-p(\bx;\x,\q)),\l{3.40}\ee
where $u$ and $p$ are given functions of the $independent$ set of
variables $\x_i$ and $\q_i$, $i=1,2,...,\nu$, and $\D_\e$ is the
normalization factor. We want to assume also that $\nu\geq1$ and we
expect that the equalities: \be
\vp(\bx,t)=u(\bx;\x(t),\q(t)),~\pi(\bx,t)=p(\bx;\x(t),\q(t))
\l{3.40a}\ee are satisfied under necessary for us choice of $\x$ and
$\q$. The fact of the matter is that Eqs. (\r{3.40a}) singles out
definite parametrization of functions $\vp(\bx,t)$ and $\pi(\bx,t)$.
Therefore, we should think that substitution of $u$ and $p$ will
transform the equalities: $$\dot\vp-\f{\d H_j(\pi,\vp)} {\d \pi}=0,~~
\dot\pi+\f{\d H_j(\pi,\vp)}{\d \vp}=0$$ into identities. This
possibility is a consequence of the fact that both differential
measures in (\r{3.40}) and (\r{3.39}) are $\d$-like.

Let us introduce for sake of clearness the lattice in the $\bx$ space
with $n$ cells. Then (\r{3.40a}) presents $n$ algebraic equalities
for $\nu$ functions of time $(\x_i(t),\q_i(t)),~i=1,2,...,\nu$ which
are independent from coordinate $\bx$.

Next, let us assume that substitution of $(\xb(t),\qb(t))$ into Eqs.
(\r{3.40a}) transform them into the identities. Notice also that
$\nu<n$. This means that the integral (\r{3.40}) is $\sim
\d_\e^{(n-\nu)}(0)\sim (1/\e)^{(n-\nu)}\to\infty$ at $\e\to0$.
Therefore, considered transformation is singular.

Considering $(\xb(t),\qb(t))$ as the unique solution of (\r{3.40a})
one can write that $\D_\e=\D_\e(\xb(t),\qb(t))$, where \be
\D_\e(\xb,\qb)=\int\prod_{i=1}^\nu\prod_{t} d\tilde\x_i d\tilde\q_i
\d_\e\le(\sum_{i=1}^\nu(u_{\xb_i}\tilde\x_i +u_{\qb_i}\tilde\q_i)\ri)
\d_\e\le(\sum_{i=1}^\nu(p_{\xb_i}\tilde\x_i
+p_{\qb_i}\tilde\q_i)\ri)\l{det}\ee since $(\xb_i(t),\qb_i(t))$ are
the necessary for us variables: the equalities \be
\sum_{i=1}^\nu(u_{\xb_i }(\bx;\xb,\qb)\tilde\x_i
+u_{\qb_i}(\bx;\xb,\qb)\tilde\q_i)=0,~~
\sum_{i=1}^\nu(p_{\xb_i}(\bx;\xb,\qb) \tilde\x_i
+p_{\qb_i}(\bx;\xb,\qb)\tilde\q_i)=0\l{**}\ee can be satisfied iff:
\be\tilde\x_i(t)=0,~\tilde\q_i(t)=0, i=1,2,...,\nu.\l{solu}\ee This
solution is unique iff all $\xb_i(t)$ and $\qb_i(t)$ are independent
even if $u$ and $p$ are not independent. Therefore, our only
requirement is the absence of the additional, hidden, equalities of
$f_\a(\xb,\qb)=0$ type, $\a=1,2...,$.

We perform the transformation (\r{1.3'}) inserting the unite
(\r{3.40}). As a result we come to the measure performing integration
over $\vp$ and $\pi$ firstly. Noting that the measures in (\r{3.40})
and in (\r{3.39}) are both $\d$-like we find: $$
DM=\prod_{t}\prod_{i=1}^\nu\f{d\x_i(t)d\q_i(t)} {\D(\xb,\qb)}
\prod_{\bx,t}\d_\e\le(\dot u-\f{\d H_j(u,p)} {\d p}\ri) \d_\e\le(\dot
p+\f{\d H_j(u,p)}{\d u}\ri)=$$\be=\prod_{t}\prod_{i=1}^\nu
\f{d\x_i(t)d\q_i(t)}
{\D(\xb,\qb)}\prod_{\bx,t}\d_\e\le(\sum_{i=1}^\nu
u_{\x_i}\dot\x_i+\sum_{i=1}^\nu u_{\q_i}\dot\q_i-\f{\d H_j(u,p)}{\d
p}\ri) \d_\e\le(\sum_{i=1}^\nu p_{\x_i}\dot\x_i+\sum_{i=1}^\nu
p_{\q_i}\dot\q_i +\f{\d H_j(u,p)}{\d u}\ri). \l{3.41}\ee Using the
auxiliary integration method \C{prepr} one can diagonalize the
arguments of last $\d$-functions in (\r{3.41}). One can write: $$
DM=\prod_t\prod_{i=1}^\nu\f{d\x_i
d\q_i}{\D(\xb,\qb)}\int\prod_t\prod_{i=1}^\nu
d\tilde{\x_i}d\tilde{\q_i}\d\le(\tilde{\x_i}- \le(\dot\x_i-\f{\pa
h_j}{\pa \q_i}\ri)\ri)\d\le(\tilde{\q_i}-\le(\dot\q_i+ \f{\pa
h_j}{\pa \x_i}\ri)\ri) \t$$\be\t\d_\e\le(\sum_{i=1}^\nu
u_{\x_i}\tilde\x_i+\sum_{i=1}^\nu u_{\q_i}\tilde\q_i +\{u,h_j\}-\f{\d
H_j(u,p)}{\d p}\ri)\d_\e\le(\sum_{i=1}^\nu
p_{\x_i}\tilde\x_i+\sum_{i=1}^\nu p_{\q_i}\tilde\q_i+\{p,h_j\}+\f{\d
H_j(u,p)}{\d u}\ri). \l{3.42}\ee It is easy to see that (\r{3.42}) is
identical to (\r{3.41}).

Let us assume now that $u(\bx;\x,\q)$, $p(\bx;\x,\q)$ and
$h_j(\x,\q)$ are chosen so that \be \{u,h_j\}=\f{\d H_j(u,p)}{\d
p},~~\{p,h_j\}=-\f{\d H_j(u,p)}{\d u},\l{3.43}\ee where Poisson
bracket
$$\{u,h_j\}=\sum_{i=1}^\nu\le\{\f{\pa u}{\pa\x_i}\f{\pa h_j}{\pa\q_i}- \f{\pa
u}{\pa\q_i}\f{\pa h_j}{\pa\x_i}\ri\}$$ and the same we should have
for $\{p,h_j\}$. Having in mind that the arguments of $\d$-functions
in (\r{3.42}) are accumulated near $\tilde\x_i=\tilde\q_i=0$ we come
to the expression: \be DM=\prod_t\prod_{i=1}^\nu\f{d\x_i
d\q_i}{\D_\e(\xb,\qb)}\d\le(\dot\x_i-\f{\pa h_j}{\pa
\q_i}\ri)\d\le(\dot\q_i+ \f{\pa h_j}{\pa \x_i}\ri) \D_\e(\x,\q),
\l{3.44}\ee where \be \D_\e(\x,\q)=\int \prod_t d\tilde\x d\tilde\q
\prod_{\bx,t}\d_\e(u_\x\tilde\x+ u_\q\tilde\q)
\d_\e(p_\x\tilde\x+p_\q\tilde\q). \l{determ}\ee have the same
structure as (\r{det}).

The Jacobian of transformation is a ratio of determinants:
$$J=\D_\e(\x,\q)/\D_\e(\xb,\qb),$$ where $(\xb,\qb)$ are
the solutions of Eqs. (\r{3.40}) and $(\x,\q)$ are the solutions of
equations \be \dot\x-\f{\pa h_j}{\pa \q}=0,~~\dot\q+ \f{\pa h_j}{\pa
\x}=0,\l{Hameq}\ee as it follows from (\r{3.44}).

It is not too hard to understand that the set of variables $(\x,\q)$
in (\r{determ}) is the same as in (\r{det}) since $u$ and $p$ must be
chosen equal to the solutions of the incident equations. Indeed,
taking into account (\r{Hameq}) and then (\r{3.43}),
$$\dot
u=\sum_{i=1}^\nu(u_{\x_i}\dot\x_i+u_{\q_i}\dot\q_i)=\{u,h_j\}=\f{\d
H_j}{\d p},~~~~\dot p=\sum_{i=1}^\nu(p_{\x_i}\dot\x_i+p_{\q_i}
\dot\q_i)=\{p,h_j\}=-\f{\d H_j}{\d u}.$$ As the result the Jacobian
of the considered transformation is equal to one, $J=1$, since the
arguments of (\r{determ}) and (\r{det}) are equal to the one of the
other.

The disappearance of $J$ leads to the absence of explicit dependence
from $\bx$. At the end one may choose $\e=0$ and turn to the
continuous $\bx$ taking $n=\infty$. As a result: \be
DM=\prod_t\prod_{i=1} ^\nu d\x_i d\q_i \d\le(\dot\x_i-\f{\pa
h_j}{\pa\q_i}\ri) \d\le(\dot\q_i + \f{\pa h_j}{\pa
\x_i}\ri).\l{3.46}\ee

The $ansatz$: \be H_j(u,p)=h_j(\x,\q)\l{3.48}\ee is natural since $u$
and $p$ must obey the incident equations. At the end, Eq. (\r{3.43})
defines the parametrization of $u$ and $p$ in terms of $\x(t)$ and
$\q(t)$ and the dynamics is defined by functional $\d$-functions in
the measure (\r{3.46}), see Eqs. (\r{Hameq}).

One can note definite conformity of considered transformation
with canonical method of Hamilton mechanics in which the
mechanical problem is divided into two parts, see e.g.
\C{arnoldd}. In our case of the field theory with symmetry the
problem also consists of two steps. First one is mapping of
$\vp(\bx,t)$, and $\pi(\bx,t)$, into factor space $W$, i.e. the
first step is the definition of functional parametrization
$u(\bx;\x,\q)$ and $p(\bx;\x,\q)$. One should assume at this
point that we must solve Eqs. (\r{3.43}) together with
(\r{3.48}) to find $u(\bx;\x,\q)$ and $p(\bx;\x,\q)$. The
second step is the dynamical problem: one must solve Eqs.
(\r{Hameq}) expanding solutions over $j(\bx,t)$. This may lead
to contradiction with an assumption that $(\x,\q)$ are $\bx$
independent quantities. The reason why the solution
$(\x(\bx,t),~\q(\bx,t))$ is impossible is shown in Appendix B.
It will be shown in the subsequent Subsection that the
dependence of $j$ on $\bx$ can not inspire the problems.

\subsection{\it Mapping into $T^*W\t C$}

We shall consider the mostly general factor space: \be W=T^*W+C,
\l{1.7l}\ee where $C$ is the zero modes manifold. Eq. (\r{1.7l})
means that we wish to consider the system which is not completely
integrable in the semiclassical limit: the conditions of
Liouville-Arnold theorem \C{arnoldd} are not hold for it and $W$ can
be compactified in that case only partly. Generally $W$ in the case
(\r{1.7l}) presents the hypertube. Its normal cross-section gives the
compact manyfold with $(\x,\q)$ coordinates. Following to the general
quantization rules only this canonical pare(s) must be quantized. It
will be shown that the remaining variables, $\la$, are $c$-numbers,
$\la\in C$ \footnote{The example from quantum mechanics:
$(angular~momentum, angle)\in T^*W$ are the $q$-numbers but
$(length~of~Runge-Lenz~ vector)\in C$ is the $c$-number in the
$H$-atom problem \C{prepr}.}. The problem of extraction of $T^*W$
subspace from $W$ can not be solved without knowledge of explicit
form of $u(\bx,t;\ga_0)$ and we would assume that $T^*W$ is not the
empty space \footnote{QED presents the example of empty $T^*W$.}.

The strictness of Eq. (\r{ab}) in addition shows how $j(\bx,t)$ must
be transformed if $\vp$ is transformed. Just this consequence allows
to define the structure of $W$. Corresponding to (\r{1.3'}) map of
$j(\bx,t)$ looks as follows: $ j(\bx,t)\to (j_\x(t),j_\q(t)),$ where
$j_\x$ and $j_\q$ are the random forces acting along axes $\x$ and
$\q$.

Following to (\r{3.48}), \be h_j(\x,\q)=h(\x,\q)+\int d\bx
u(\bx;\x,\q) j(\bx,t).\l{k14}\ee Therefore
$$ DM=\prod_{t}\prod_{i} d\x_i(t) d\q_i(t)\d\le(\dot\x_i(t)-
\o_{\q_i}(\x,\q)-\int d\bx \f{\pa
u(\bx,\x,\q)}{\pa\q_i(t)}j(\bx,t)\ri)\t$$\be\t\d\le
(\dot\q_i(t)+\o_{\x_i}(\x,\q)+ \int d\bx\f{\pa
u(\bx,\x,\q)}{\pa\x_i(t)}j(\bx,t)\ri),\l{k15}\ee where \be
\o_{X_i}=\f{\pa h}{\pa X_i},~X=(\x,\q),\l{k16}\ee is the "speed" in
the $W$ space.

One may simplify calculations using the equality \C{prepr}:
$$ \d\le(\dot\x(t)- \o_\q-\int d\bx \f{\pa
u(\bx,\x,\q)}{\pa\q(t)}j(\bx,t)\ri)=\lim_{j_\x=e_\x=0} e^{-(i/2)\int
dt\f{\d}{\d j_{\x}(t)}\f{\d}{\d e_{\x}(t)}}\t$$\be\t e^{2i\int dt
e_{\x}(t)\int d\bx \f{\pa u(\bx,\x,\q)}{\pa\q(t)}j(\bx,t)}
\d\le(\dot\x_i(t)- \o_\q-j_{\x_i}(t)\ri).\l{k17}\ee This equality
follows from Fourier transformation of $\d$-function. The same
transformation of argument of second $\d$-function in (\r{k15}) can
be done.

Inserting (\r{k17}) into the expression for $\R_{mn}$ the action of
operator (\r{2.6b}) gives new perturbations generating operator \be
2\bk(j,e)=\int dt\le\{\f{\d}{\d j_\x}\cdot\f{\d}{\d e_\x}+ \f{\d}{\d
j_\q}\cdot\f{\d}{\d e_\q} \ri\}\l{k18}\ee and new auxiliary field \be
e_c(\bx,t)=\le\{e_{\q}(t)\cdot\f{\pa u(\bx;\x,\q)}{\pa\x(t)} -
e_{\x}(t)\cdot\f{\pa u(\bx;\x,\q)}{\pa\q(t)}\ri\}. \l{k19}\ee At the
very end of calculations one must take all $j_{\x},~j_\q$ and
$e_\x,~e_\q$ equal to zero. The transformed measure looks as follows:
\be DM=\prod_{t}\prod_{i=1}^\nu d\x_id\q_i\d(\dot\x_i(t)- \o_\q
-j_{\x_i}(t))\d(\dot\q_i(t)+\o_\x+ j_{\q_i}(t)),\l{k15x}\ee where new
forces, $j_{\x}(t),j_\q(t)$ are $\bx$ independent. Eqs. (\r{k18}),
(\r{k19}), (\r{k15x}) and \be
\R(z)=\lim_{\{\x,\q;e_\x,e_\q\}=0}e^{-i\bk(j,e)}\int DM
e^{-iU(u,e_c)} e^{-N(z; u)} \l{2.13x}\ee form the transformed theory
in which each degree of freedom is excited by individual source,
$j_{\x_i}$ and $j_{\q_i}$ and the $\bx$ dependence have been
integrated.

Introduction of $(j_\x(t),j_\q(t))$ ends the mapping of quantum
theory into the linear space $W$. The latter means that $W$ is
an isotropic and homogeneous \C{smalee} and as the result the
perturbation theory in $W$ is extremely simple. That is why the
problem of $\R_{mn}$ in $W$ becomes calculable in all orders of
$\hbar$ even in the case $u(\bx,t)\neq$ const. It must be noted
also that (\r{2.13x}) presents the expansion over $\hbar^2$~
\C{prepr}. This is readily seen from the estimation:
$U/\hbar=O(\hbar^2)$

It is quite possible that not all parameters $\{\ga\}\in W$ are
$q$-numbers. To define the structure of factor space $W$ one
must extract from $\{\ga\}$ the set of the canonically
conjugated pares. We leave for them the same notations $\x$ and
$\q$. This set will form simplectic subspace, $\{\x,\q\}\in
T^*W$. Through $\la$ we will denote other coordinates, $\la\in
C$. It is suitable to introduce the conjugate to $\la$ the
auxiliary variables $\a$: $$ DM=\prod_{t} d^\nu\x(t)d^\nu\q
\d^{(\nu)}(\dot\x(t)- \o_\q
-j_{\x}(t))\d^{(\nu)}(\dot\q(t)+\o_\x+ j_{\q_i}(t))\t$$$$\t
\prod d\la(t)d\a \d(\dot\la(t)- \o_\a
-j_{\la}(t))\d(\dot\a(t)+\o_\la+ j_{\a}(t))$$ and $$
2\bk(j,e)=\int dt\le\{\f{\d}{\d j_\x}\cdot\f{\d}{\d e_\x}+
\f{\d}{\d j_\q}\cdot\f{\d}{\d e_\q}+\f{\d}{\d
j_\la}\cdot\f{\d}{\d e_\la}+\f{\d}{\d j_\a}\cdot\f{\d}{\d e_\a}
\ri\},$$ $$ e_c(\bx,t)=\le\{e_{\q}(t)\cdot\f{\pa u}{\pa\x(t)} -
e_{\x}(t) \cdot\f{\pa u}{\pa\q(t)}\ri\}+\le\{e_{\a}\cdot\f{\pa
u}{\pa\la(t)}- e_{\la}\cdot\f{\pa u}{\pa\a(t)}\ri\}$$ to search
the consequences of such enlargement assuming that $u$ does not
depend on $\a$: \be \f{\pa u}{\pa\a}\approx 0.\l{2*}\ee We want
to show that only the canonically conjugate pares quantize.
Having (\r{2*}) $e_c$ looks as follows: \be
e_c(\bx,t)=\le\{e_{\q}(t)\cdot\f{\pa u}{\pa\x(t)} - e_{\x}(t)
\cdot\f{\pa u}{\pa\q(t)}\ri\}+e_{\a}\cdot\f{\pa u}{\pa\la(t)}
\l{3*}\ee and $$ DM=\prod_{t} d^\nu\x(t)d^\nu\q
\d^{(\nu)}(\dot\x(t)- \o_\q
-j_{\x}(t))\d^{(\nu)}(\dot\q(t)+\o_\x+ j_{\q_i}(t))\t$$$$\t
\prod d\la(t)d\a(t) \d(\dot\la(t)
-j_{\la}(t))\d(\dot\a(t)+\o_\la+ j_{\a}(t)),$$ where the
conditions (\r{2*}) were taken into account. Therefore,
dependence on $e_\la$ disappears in $e_c$, i.e. \be
2\bk(j,e)=\int dt\le\{\f{\d}{\d j_\x}\cdot\f{\d}{\d e_\x}+
\f{\d}{\d j_\q}\cdot\f{\d}{\d e_\q}+\f{\d}{\d
j_\a}\cdot\f{\d}{\d e_\a} \ri\}.\l{c1}\ee since all derivatives
over $e_\la$ are equal to zero. Next, as it follows from
(\r{c1}), the derivatives over $j_\la$ also disappears in
$\bk$. For this reason we can put $j_\la=0$:
$$ DM=\prod_{t} d^\nu\x(t)d^\nu\q \d^{(\nu)}(\dot\x(t)- \o_\q
-j_{\x}(t))\d^{(\nu)}(\dot\q(t)+\o_\x+ j_{\q_i}(t))\t$$$$\t \prod
d\la(t)\d(\dot\la(t) )d\a(t) \d(\dot\a(t)+\o_\la+ j_{\a}(t)),$$
Remembering (\r{2*}) one may perform the shift:
$\dot\a\to\dot\a-\o_\la-j_\a$. As a result: $$ DM=d\la(0)\prod_{t}
d^\nu\x(t)d^\nu\q \d^{(\nu)}(\dot\x(t)- \o_\q
-j_{\x}(t))\d^{(\nu)}(\dot\q(t)+\o_\x+ j_{\q_i}(t)),$$ where the
integral over $\a$ was omitted and the definition: $$\int \prod_t
d\la(t)\d(\dot\la(t))=\int d\la(0)$$ was used.

Therefore, the formalism naturally extracts the set of $q$-numbers
and defines the measure of integrals over $c$-numbers.

Let us introduce the coordinates $\q$ through the condition: \be
h=h(\q).\l{}\ee Then equation of motion in $T^*W$ space looks as
follows: \be \dot\x=\f{\pa h_j}{\pa\q}=\o(\q)+j_\x,~~ \dot\q=-\f{\pa
h_j}{\pa\x}=j_\q,\l{equa}\ee i.e. $\x$ can be considered as the
generalized coordinate and $\q$ is the conserved in the semiclassical
approximation canonically conjugate generalized momentum, when
$j_\x=j_\q=0$. The Eqs. (\r{equa}) have following exact solutions:
$$\q_j(t)=\q_0+\int_{0}^{+\infty} dt'g(t-t')j_\q(t') \equiv
\q_0+\q(t;j),$$ \be \x_j(t)=\x_0+ \int_{0}^{+\infty}
dt'g(t-t')(\o(\q_0+\q)+j_\x(t'))\equiv
\x_0+\o(t;\q_0+\q)+\x(t;j),\l{grinf}\ee where the boundary
conditions: \be \x(0)=\x_0,~~\q(0)=\q_0 \l{bcon}\ee were applied. So
\be \lim_{j=0}\q_j=\q_0,~~\lim_{j=0}\x_j= \o_0+\o(\q_0)t. \l{bcon
2}\ee The Green function $g(t-t')$ has the extremely simple form
\C{prepr}: \be g(t-t')=\Th(t-t'),~~ \Th(0)=1.\l{green}\ee This
explains why the mapping into $W$ space is useful. Notice that the
singularity of $g(t-t')$ is integrable.

\section{Mass-shell particle production}

The result of integration over $\x(t)$ and $\q(t)$ looks as follows:
\be \R_{mn}(z) =\lim_{\x=\q=e_\x=e_\q=0} e^{-i\bk(je)} \int dM
e^{iU(u,e_c)}N(z;u)^mN^*(z;u)^n,\l{2.14} \ee where $$ dM=d\la d\x_0
d\q_0. $$ We will consider the simplest case: $\dim T^*W=2$ and \be
2\bk=\int_{-\infty}^{+\infty} dt\le\{\f{\d}{\d\x(t)} \Th(t-t')
\f{\d}{\d e_\x(t')} \ri. +\le. \f{\d}{\d\q(t)} \Th(t-t') \f{\d}{\d
e_\q(t')} \ri\}dt'. \l{2.15} \ee The functional $U$ can be written in
the general form: \be U(u,e_c)=\int d\bx d\tau
(e_c^3(\bx,\tau)u(\bx;\x(\tau),\q(\tau))+...),\l{U1}\ee where the
dots signify higher orders over $e_c^{2r+1},~r=2,3,...$  The
auxiliary variable $e_c$: $$e_c(\bx,t)=\le\{e_{\q}(t)\cdot\f{\pa
u(\bx;\x,\q)}{\pa\x(t)} - e_{\x}(t)\cdot\f{\pa u(\bx;\x,\q)}
{\pa\q(t)}\ri\}$$ was defined in (\r{k19}) and \be
u(\bx;\x,\q)=u(\bx;\x_0+\o(t;\q_0+\q)+\x(t), \q_0+\q(t)).\l{u=}\ee
Notice that the integration in (\r{2.15}) is performed along real
time axis. This becomes possible if $u(x)$ is the regular function.
Otherwise we must conserve definition of theory on the complex time
plane until the very end of calculations.

By definition $U$ must be the odd function of $e_c$, see
(\r{U1}) and (\r{Uex}). This generates following lowest over
$U$ term: \be \sim\bk^3 U(u,e_c) N(z;u)^mN^*(z;u)^n\l{3.2}\ee
and the common term of our perturbation theory is: \be
\sim\bk^{3l} U(u,e_c)^l N(z;u)^m
N^*(z;u)^n=\bk^{3l}O(\Ga^{2(n+m)}) \l{3.2a}\ee since $$
N(z,u)=\int d\o_1(z,q) \Ga(q;u)\Ga^*(q;u),$$ see (\r{N}), where
\be \Ga(q;u)=\int dx e^{-ixq} \pa^2u(x)=\int dx \pa_\mu(\pa^\mu
+2iq^\mu)\le\{e^{-iqx}\vp(x)\ri\}, \l{ga2}\ee see (\r{ga1}).

Notice that \be \lim_{\x=\q=0}\Ga(q;u)=0 \l{lim}\ee because of the
condition (\r{ga1}). We will consider the fields:  \be
\lim_{t\to\pm\infty}u(\bx,t; \x_0,\q_0)=\lim_{t\to\pm\infty}\pa_t
u(\bx,t; \x_0,\q_0)=0,~~\forall(\x_0,\q_0),\l{der3}\ee assuming that
this condition is rightful in the infinitesimal neighborhoods of
$\x_0$ and $\q_0$.

The variational derivative over $\x(t')$ gives:
$$ \f{\d}{\d\x(t')}\lim_{t\to\pm\infty}(\pa_t
+2iq_0)\le.\le\{e^{-iqx}u(\bx;\x(t),
\q(t))\ri\}\ri|_{\x=\x_0,\q=\q_0}=$$\be=\lim_{t\to\pm\infty}(\pa_t
+2iq_0)\le\{e^{-i(q_0t-\bq\bx)}\f{\pa}{\pa\x_0}u(\bx,t;\x_0,
\q_0)\ri\}\d(t-t')\l{der1}\ee since the derivative is calculated in
the vicinity $\x_0$. The same we will have for higher derivatives:
$$ \prod_{i=1}^k\f{\d}{\d\x(t_i')}\prod_{j=1}^l\f{\d}{\d\x(t_j')}
\lim_{t\to\pm\infty}(\pa_t +2iq_0)\le\{e^{-iqx}u(\bx;\x(t),
\q(t))\ri\}=$$\be=\lim_{t\to\pm\infty}(\pa_t +2iq_0)
\le.\le\{e^{-i(q_0t-\bq\bx)}\f{\pa^k}{\pa\x_0^k}
\f{\pa^l}{\pa\q_0^l}u(\bx,t;\x_0,
\q_0)\ri\}\ri|_{\x=\x_0,\q=\q_0}\prod_{i=1}^k\d(t-t_i')
\prod_{j=1}^l\d(t-t_j').\l{der2}\ee Integration over $t_i'$ and
$t_j'$ reduces $\d$-functions into $\th$-functions and last ones may
restrict the range of integration over time variables $\tau$ of the
convergent integrals, see Eq. (\r{U1}) and Appendix A. Therefore, the
asymptotic over $t$ is defined only by derivatives of $u(\bx,t;
\x_0,\q_0)$ over $\x_0$ and $\q_0$.

It is shown in Appendix A that if (\r{der3}) is rightful then the
same must exist for all derivatives over $\x_0$ and $\q_0$. This ends
the prove of Eq. (\r{o2}).

\section{Concluding remarks}

--- It is important to have in mind that the transformation (\r{1.3'})
is singular, i.e. the inverse to (\r{1.3'}) transformation is
impossible. The latter is significant for self-consistence of the
approach: Eq. (\r{o2}) means that the generated by $u(\bx,t;\ga_0)$
constraints are so important that even a notion of plane waves is
lost in the theory, or, in other words, Eq. (\r{o2}) means that the
fluctuations of $\ga(t)$ compose a complete set of contributions and
there is no need to take into account other ones.

--- Our general result, Eq. (\r{o2}), can be extended
on gluon production considering Yang-Mills theory as the theory
with symmetry. But Gribov ambiguity \C{gribovv} prevents
proving of Eq. (\r{o2}) for non-Abelian gauge theory canonical
formalism if $u\neq0$. It can be shown at the same time that
GCP based formalism gives in each order over $\hbar$ the gauge
invariant terms \footnote{Since the gauge invariant quantity,
$\R_{mn}$, is calculated} and for this reason there is no
necessity to extract gauge degrees of freedom in it. The
tentative consideration of that solution was given in
\C{jmp-grib} and the complete description will be published
later.

--- Transformed perturbation theory presents the expansion over
$\hbar^2$, i.e. it is not the WKB expansion, see Appendix A. A short
discussion of the structure of new perturbation series is given in
\C{prepr}.

A few remarks concerning unsolved problems at the end of the paper.

--- There exists two ways to compute $\R_{mn}$ having
the non-trivial $u(x)$. First one was described at the end of Sec.2
and the GCP based formalism is given in Sec.3. One can think that
both methods must lead to the same result (\r{o2}) since the primary
formula (\r{2.13ax}) is the same for both approaches. But I can not
prove this equivalence because of extremal complexity of the first
approach. It is possible that the problem is connected with
transparent mechanism of accounting of the symmetry constraints in
the canonical formalism. Notice that the mapping into the simplectic
space $T^*W$ is the one of possible ways to realize Dirac's
\C{diracc} programm.

--- It must be noted that if $u(x)$ have finite energy then GCP
formulas are applicable at all distances and does not require
infrared dimensional parameter $\La$. It is not clear for this reason
how to join GCP approach with canonical formalism.

--- There exists the problem with interaction at small distances
where the perturbative QCD formalism is presumably strict. For
example, it is unclear how to explain the "asymptotic freedom" effect
in the GCP formalism since it is impossible to introduce the "running
coupling constant" in the GCP strong coupling perturbation theory
over inverse interaction constant, see the example in Appendix A,
without divergences and without even notion of "gluon".

--- The enlargement of the GCP approach on the non-Abelian
gauge theories assumes presence of the quark fields. This will be
possible if the quark fields contribution is the {\it invariant} of
the factor group $G/G_w$ \C{prepr} since only in this case the fields
of quark sector do not give an influence on the vector fields.

--- By all appearance, if the {\it unitary} $S$-matrix exists in the
general relativity then even the notion of "graviton" disappeared in
this theory. In other words, the quantum perturbations must be
described in terms of the fluctuations of metric, $u_{\mu\nu}$, under
the conditione that $u_{\mu\nu}\in W$ since the general relativity
symmetry constraints {\it must} be taken into account. The question
of {\it singular} metric demands separate consideration.

I hope to look into some of this questions in the subsequent
publications.

\renewcommand{\theequation}{a.\arabic{equation}}
\section {Appendix A. Example of massless $\vp^4$ theory}\0

Let us consider \be \R_{10}=\lim_{\x=\q=e_\x=e_\q=0} e^{-i\bk(je)}
\int dM e^{iU(u,e_c)}N(u), \l{ap1}\ee where $N(u)=N(z;u)|_{z=1}$, \be
N(u)=\int \f{d\bq}{(2\pi)^3q_0}
\Ga(q;u)\Ga^*(q;u),~q_0=+\sqrt{\bq^2}\l{ap2}\ee with \be
\Ga(q;u)=\int d\bx\int_{-\infty}^{+\infty} dt \pa_t\le[e^{-iqx}(\pa_t
+iq)u(\bx;\x_j,\q_j)\ri]\l{ap3}\ee equal to zero if $\x_j(t)=\x_0$
and $\q_j(t)=\q_0$, i.e. \be \Ga(q;u)|_{j=0}=0 \l{ap'}\ee The
operator \be 2\bk(j,e)=\int_{-\infty}^{+\infty} dt\le\{\f{\d}{\d
j_\x(t)} \f{\d}{\d e_\x(t)} + \f{\d}{\d j_\q(t)} \f{\d}{\d e_\q(t)}
\ri\} \l{ap4}\ee and \be e_c(\bx,t;\x_j,\q_j)=\le\{e_{\q}(t)\f{\pa
u(\bx;\x_j(t),\q_j(t))}{\pa\x_0} - e_{\x}(t)\f{\pa
u(\bx;\x_j(t),\q_j(t))}{\pa\q_0}\ri\} \l{ap5}\ee

To generate perturbation series one should expand the operator:
$$e^{-i\bk(je)}=\sum_{n_\x,n_\q=0}^{\infty}\f{(-i/2)^{n_\x+n_\q}}{n_\x!n_\q!}
\int_{-\infty}^{+\infty}\prod_{l_\x=1}^{n_\x}dt_{l_\x}\f{\d}{\d
j_\x(t_{l_\x})} \f{\d}{\d e_\x(t_{l_\x})}
\prod_{l_\q=1}^{n_\q}dt'_{l_\q}\f{\d} {\d j_\q(t'_{l_\q})}\f{\d}{\d
e_\q(t'_{l_\q})}$$ Let us consider now the expansion: $$
e^{iU(u,e_c)}= \sum_{n_\x,n_\q=0}^{\infty}\int_{-\infty}^{+\infty}
\prod_{k_\x=1}^{n_\x} dt_{k_\x} e_\x(t_{k_\x})\prod_{k_\q=1}^{n_\q}
dt'_{k_\q}
e_\x(t'_{k_\q})C_{n_\x,n_\q}(u;t_1,...,t_{n_\x},t'_1,...,t'_{n_\q}),
$$ where part of $C_{n_\x,n_\q}$ may be equal to zero. Therefore, $$
\lim_{e_\x=e_\q=0}e^{-i\bk(je)}e^{iU(u,e_c)}=\sum_{n_\x,n_\q=0}^{\infty}
\int_{-\infty}^{+\infty}\prod_{k_\x=1}^{n_\x} dt_{k_\x}\f{-i\d}{2\d
j_\x(t_{l_\x})} \prod_{k_\x=1}^{n_\q} dt_{k_\q}\f{-i\d}{2\d
j_\q(t_{l_\q})}\t$$$$\t
C_{n_\x,n_\q}(u;t_1,...,t_{n_\x},t'_1,...,t'_{n_\q})=\hat O
e^{iU(u,\hat e_c)},$$ where \be 2i\hat e_c=\le\{\f{\d}{\d
j_\q(t)}\f{\pa u(\bx;\x_j(t),\q_j(t))}{\pa\x_0} - \f{\d}{\d
j_\x(t)}\f{\pa u(\bx;\x_j(t),\q_j(t))}{\pa\q_0}\ri\}.\l{ap}\ee

As the result, one can rewrite (\r{ap1}) in the form: \be
\R_{10}=\lim_{\x=\q=0} \int dM \hat O e^{iU(u,\hat e_c)}N(u),
\l{ap6}\ee where $\hat O$ means that the derivatives should stay to
the left of all function on which it act. Considering the model
(\r{lagr}) one can find $u\sim g^{-1/2}$, see Eq. (\r{equ}), and \be
U(u,\hat e_c)=g\int d\bx dt \hat e_c^3 u .\l{exa}\ee Therefore,
expansion over $U(u,\hat e_c)$ gives series over $1/g$. Taking into
account (\r{ap3}) it is easy to see that the lowest order gives the
term $\sim U(u,\hat e_c)^2$. Next, one can find that $\hat e_c\sim
\hbar$ in the units of $\hbar$. Therefore the expansion over
$(U/\hbar)$ generates series over $\hbar^2$. Notice also that each
order over $\hbar^2$ is real, see (\r{ap6}).

Noting that $N=O(\Ga^2)$ and taking into account comment to (\r{ap2})
one can find inserting (\r{exa}) into (\r{ap6}) that the lowest
nonequal to zero contribution looks as follows:
$$ \R_{10}=\lim_{j_\x=j_\q=0} \int dM \hat O U(u,\hat e_c) N(u)
+...=$$$$ =\lim_{j_\x=j_\q=0} \int dM \int
d\bx_1\int_{-\infty}^{+\infty} dt_1\hat O
[\hat{e}_c^3(\bx_1,t_1;\x_j,\q_j) u(\bx_1;\x_j,\q_j)_{t_1} ]
N(u)+...$$ Let as consider for the sake of simplicity action of the
first term in (\r{ap}). Then:
$$ \R^{(1)}_{10}=\lim_{j_\x=j_\q=0} \int dM\int \f{d\bq_1}{(2\pi)^3q_{10}}
\int d\bx_1\int_{-\infty}^{+\infty}
dt_1 \hat O\t$$$$\t \le[ \f{\d}{\d j_\q(t_1)}\f{\pa
u(\bx_1;\x_j,\q_j)_{t_1}}{\pa\x_0}\ri]^3u(\bx_1;\x_j,\q_j)_{t_1}
\t$$\be\t \Ga(q_1;u)\Ga^*(q_1;u)+...,\l{ap8}\ee where the
differential operators act on all right standing functions of $u$.

Taking into account the definition of $\Ga$'s in (\r{ap3}) we should
be interested just in the results of action of differential operators
$\d/\d j_\q(t_1)$:
$$\lim_{j=0}\f{\d}{\d j_\q(t_1)}\Ga(u)=\lim_{j=0} \int
d\bx\int_{-\infty}^{+\infty} dt' \pa_{t'}\le[e^{-iqx}(\pa_{t'}
+iq_0)\f{\d}{\d j_\q(t_k)}u(\bx,t';\x_j,\q_j)\ri]=$$$$=\lim_{j=0}
\int d\bx \int_{-\infty}^{+\infty} dt' \pa_{t'}\le[e^{-iqx}(\pa_{t'}
+iq_0)\f{\pa u(\bx,t';\x_0,\q_0)} {\pa\q_0}\f{\d \q_j(t')}{\d
j_\q(t_1)}\ri].$$ Noting that
$$\q_j(t')=\q_0+ \int_{0}^{+\infty} dt''g(t'-t'')j_\q(t'')$$ we will have:
$$ \f{\d \q_j(t')}{\d j_\q(t_1)}=\int_{-\infty}^{+\infty} dt''
\Th(t'')g(t'-t'')\d(t''-t_1))=\Th(t_1)g(t'-t_1) $$

Therefore, \be\lim_{j=0}\f{\d\Ga(u)}{\d j_\q(t_1)}=\int d\bx'
\int_{-\infty}^{+\infty} dt'
\pa_{t'}\le[e^{-i(q_0t'-\bq\bx')}(\pa_{t'} +iq_0)\f{\pa
u(\bx',t';\x'_0,\q_0)} {\pa\q_0}\Th(t_1)\Th(t'-t_1)\ri]\l{ap7}\ee
since $g(t-t')=\Th(t-t')$. Using this result one nontrivial term in
$\R_{10}$ looks as follows:
$$ \R^{(1)}_{10}=\lim_{j_\x=j_\q=0} \int dM\int
\f{d\bq_1}{(2\pi)^3q_{10}} \int d\bx_1\int_{-\infty}^{+\infty}
dt_1\t$$\be\t \f{\d\Ga(q_1;u)}{\d j_\q(t_1)}\f{\d\Ga^*(q_1;u)}{\d
j_\q(t_1)}  \f{\d}{\d j_\q(t_1)}\le\{\le[\f{\pa
u(\bx_1;\x_j,\q_j)_{t_1}}{\pa\x_0}\ri]^3u(\bx_1;\x_j,\q_j)_{t_1}
\ri\}+...,\l{ap9}\ee where the higher derivatives of $\Ga$ also were
not shown for the sake of simplicity.

As it follows from (\r{ap7}) the derivatives of $\Ga$'s are
proportional to $\Th$-functions which restricts the range of
integration over $t_1$ and $t_2$. One can rewrite (\r{ap9}) in the
form: $$ \R^{(1)}_{10}=\lim_{j_\x=j_\q=0} \int dM\int
\f{d\bq_1}{(2\pi)^3q_{10}} \int \prod_k
d\bx'_k\int_{-\infty}^{+\infty} \prod_l dt'_l \t$$$$\t
\pa_{t_1'}\pa_{t_2'}\le\{\le[e^{-i(q_0t_1'-\bq\bx_1')}(\pa_{t_1'}
+iq_0)\f{\pa u(\bx_1',t_1';\x'_0,\q_0)} {\pa\q_0}\ri]\ri.\t$$$$\t\le.
\le[e^{-i(q_0t_2'-\bq\bx_2')}(\pa_{t_2'} +iq_0)\f{\pa
u(\bx_2',t_2';\x'_0,\q_0)}
{\pa\q_0}\ri]\ri.\t$$$$\t\le.\int_{-\infty}^{+\infty} dt_1
\Th(t_1)\Th(t'_1-t_1)\Th(t'_2-t_1)\f{\d}{\d j_\q(t_1)}\le(\le[\f{\pa
u(\bx_1;\x_j,\q_j)_{t_1}}{\pa\x_0}\ri]^3u(\bx_1;\x_j,\q_j)_{t_1}
\ri)\ri\}+...$$ Only the typical term was shown here. Therefore, we
should investigate \be \lim_{t'_1,t'_2\to\pm\infty}\f{\pa
u(\bx_1',t_1';\x'_0,\q_0)}{\pa\q_0} \f{\pa u(\bx_2',t_2';\x'_0,\q_0)}
{\pa\q_0} \l{}\ee times the function which is finite in this limits,
i.e. if this limits are equal to zero then $\R_{10}$ is also equal to
zero.

It is easy to see that if (\r{der3}) is rightful then \be
\lim_{t\to\pm\infty}\f{\pa^k}{\pa\x_0^k}
\f{\pa^l}{\pa\q_0^l}u(\bx,t;\x_0, \q_0)=0.\l{der4}\ee Indeed, one can
consider the expansion: $$u(\bx,t;\x_0+\ve_\x,\q_0+\ve_\q)=
\sum_{n_\x,n_\q=0}^\infty \f{\ve_\x^{n_\x}\ve_\q^{n_\q}}
{n_\x!n_\q!}\f{\pa^n_\x}{\pa\x_0}\f{\pa^n_\q}{\pa\q_0}
u(\bx,t;\x_0,\q_0)$$ for infinitesimal $\ve_\x,~\ve_\q$. Therefore,
if (\r{der3}) is rightful for all $(\x_0, \q_0)$ then (\r{der4}) is
also rightful since $\ve_\x,~\ve_\q$ are arbitrary. This proves
(\r{o2}) in all orders over $\hbar^2$.

\renewcommand{\theequation}{b.\arabic{equation}}
\section{Appendix B. Space-time local transformation}\0

Let us consider the case: $\x=\x(\bx,t)$ and $\q=\q(\bx,t)$. In this
case one must insert the unit: \be 1=\f{1}{\D}\int D\x D\q
\prod_{\bx,t}\d(\vp(\bx,t)- u(\bx,\x(\bx,t),\q(\bx,t)))
\d(\pi(\bx,t)- p(\bx,\x(\bx,t),\q(\bx,t))),\l{y1}\ee where $\D$ is
the normalization factor, \be D\x D\q=\prod_{\bx,t}\prod_i^{\nu}
d\x_i(\bx,t) d\q_i(\bx,t) \l{y2}\ee and $(u,p)$ are given functions
of $\bx$ and $(\x(\bx,t), \q(\bx,t))$. The "Hamiltonian" has the same
form.

If the solution of equations: \be \vp(\bx,t)=
u(\bx;\x(\bx,t),\q(\bx,t)),~~\pi(\bx,t)=p(\bx;\x(\bx,t),\q(\bx,t))
\l{y3}\ee is $\xb(\bx,t),\qb(\bx,t)$ then, see (\r{det}):
$$\D(\xb,\qb)=\int D\tilde\x D\tilde\q \prod_{\bx,t}
\d\le(\sum_{i}^\nu\le\{u_{\xb_i}(\bx;\xb,\qb)\tilde\x_i(\bx,t)+
u_{\qb_i}(\bx;\xb,\qb)\tilde\q_i(\bx,t)\ri\}\ri)\t$$
\be\t\d\le(\sum_{i}^\nu\le\{\pi_{\xb_i}(\bx;\xb,\qb)\tilde\x_i(\bx,t)+
\pi_{\qb_i}(\bx;\xb,\qb)\tilde\q_i(\bx,t) \ri\}\ri).\l{y4}\ee
We should have again \be \D^{-1}(\xb,\qb)\neq0.\l{y5}\ee

Using the method of auxiliary integration one come to the expression:
$$ DM=\prod_{\bx,t}
\prod_{i}^{\nu} d\x_i d\q_i \d\le(\dot\x_i- \f{\d
h_j}{\d\q_k}\ri)\d\le(\dot\q_i+ \f{\d h_j}{\d\x_k}\ri)\t$$$$\t
\f{1}{\D(\xb,\qb)}\int D\tilde\x D\tilde\q\prod_{\bx,t} \d
\le(\sum_{i}^\nu\le\{u_{\x_i}(\bx;\x,\q)\tilde\x_i(t)+
u_{\q_i}(\bx;\x,\q)\tilde\q_i(t)\ri\}\ri)\t$$
\be\t\d\le(\sum_{i}^\nu\le\{\pi_{\x_i}(\bx;\x,\q)\tilde\x_i(t)+
\pi_{\q_i}(\bx;\x,\q)\tilde\q_i(t) \ri\}\ri),\l{y6}\ee if the
equations:
$$u_{\xb_i}(\bx;\xb,\qb)\tilde\x_i(\bx,t)=-
u_{\qb_i}(\bx;\xb,\qb)\tilde\q_i(\bx,t),~\pi_{\xb_i}(\bx;\xb,\qb)
\tilde\x_i(\bx,t)=-\pi_{\qb_i}(\bx;\xb,\qb) \tilde\q_i(\bx,t)$$ have
the unique solution $$\tilde\x_i(\bx,t)=\tilde\q_i(\bx,t)=0.$$ Let us
assume that this conditions are satisfied. The ratio of determinants
is again canceled for the same reasons as in (\r{3.44})

At the same time we must have: \be \{u(\bx,\x,\q),h_j\}-\f{\d H_j}{\d
p(\bx,\x,\q)}=0,~~\{p(\bx,\x,\q),h_j\}+\f{\d H_j}{\d u(\bx,\x,\q)}=0,
\l{y7}\ee where the Poisson bracket:
$$\{u(\bx,\la),h_j\}=\f{\pa u(\bx,\x(\bx,t,),\q(\bx,t))}{\pa\x
(\bx,t,)}\f{\d h_j}{\d\q(\bx,t) }-\f{\pa
u(\bx,\x(\bx,t,),\q(\bx,t))}{\pa\q(\bx,t,)}\f{\d h_j}{\d\x(\bx,t)}
$$ and the same for bracket $\{p(\bx,\la),h_j\}$. Next, the Eqs. (\r{y7})
together with the same equality for $\{p(\bx,\la),h_j\}$ lead to the
equal space-time Poisson equations: \be
\{u(\bx,\x(\bx,t),\q(\bx,t)),u(\bx,\x(\bx,t),\q(\bx,t))\}=
\{p(\bx,\x,\q),p(\bx,\x,\q)\}=0 \l{y8}\ee and
\be\{u(\bx;\x(\bx,t),\q(\bx,t)),p(\bx;\x(\bx,t),\q(\bx,t))\}=1\l{y9}\ee
if the $ansatz$ (\r{3.48}) is taken into account. The last equality
can not be satisfied since $u(\bx,\x,\q)$ and $p(\bx,\x,\q)$ are not
the independent quantities.

\begin{acknowledgements}
The GCP based formalism was reported many times and I was
trying to keep in mind all critical comments gratefully.
Present paper is the response on the valid criticism of P.
Kulish.
\end{acknowledgements}

\theendnotes

\end{document}